\documentclass[usenatbib,referee]{mn2e}

\usepackage{graphicx}
\usepackage{rotating}

\newcommand{\lsol}{\hbox{$\rm L_\odot$}}
\newcommand{\msol}{\hbox{\,$\rm M_\odot$}}

\newcommand{\e}[1]{$10^{#1}$}
\newcommand{\ee}[1]{$\times 10^{#1}$}         
\newcommand{\kms}{~km\,s$^{-1}$} 
\newcommand{\cm}[1]{~cm$^{#1}$}
\newcommand{\microm}{\,$\mu$m}
\newcommand{\VLSR}{V$_{\rm LSR}$}

\newcommand{\co}{$^{12}$CO}    
\newcommand{\oh}{OH(1720 MHz)}

\newcommand{\brg}{Br\,${\gamma}$}
\newcommand{\hei}{He\,{\sc i}}
\newcommand{\h}{H$_2$}
\newcommand{\hi}{H\,{\sc i}}
\newcommand{\hr}{H\,{\sc ii}}

\newcommand{\aap}{A\&A}

\newcommand{\aj}{AJ}
\newcommand{\apj}{ApJ}
\newcommand{\apjl}{ApJL}
\newcommand{\apjs}{ApJS}
\newcommand{\mnras}{MNRAS}

\newcommand{\pasp}{PASP}
\newcommand{\nat}{Nature}

\begin{document}

\title[The Eye of the Tornado]
{The Eye of the Tornado---an isolated, high mass young stellar object
near the Galactic centre}

\author[Burton et\ al.]
       {M. G. Burton$^1$, J. S. Lazendic$^{2,3}$\thanks{Current
       address: Harvard-Smithsonian Center for Astrophysics, 60 Garden
       Street, Cambridge, MA 02138, USA}, F. Yusef-Zadeh$^{4}$,
       M. Wardle$^5$ \\ $^1$ School of Physics, University of New
       South Wales, Sydney NSW 2052, Australia\\ $^2$ School of
       Physics A28, University of Sydney, Sydney NSW 2006, Australia\\
       $^3$ Australia Telescope National Facility, CSIRO, PO Box 76,
       Epping NSW 1710, Australia \\ $^4$ Department of Physics and
       Astronomy, Northwestern University, Evanston, IL 60208, USA \\
       $^5$ Department of Physics, Macquarie University, Sydney, NSW
       2019, Australia}

\date{ }


\maketitle


\begin{abstract}
We present infrared (AAT, UKIRT) and radio (VLA, SEST) observations of
the Eye of the Tornado, a compact source apparently near the head of
the Tornado Nebula. The near-infrared
\brg\ and \hei\ lines are broad (FWHM $40$ and $30$\,\kms, respectively) 
and have a line centre at V$_{\rm LSR} \sim -$205\kms.  This
corresponds to a feature at the same velocity in the $\rm
^{12}CO$\,J=1--0 line profile.  The kinematic velocity derived from
Galactic rotation places the Eye at the distance of the Galactic
Centre (i.e.\ 8.5\,kpc) and separated (probably foreground) from the
Tornado Nebula. Four knots of emission are seen in the \brg\ line and
at 6 and 20\,cm.  Together with the flat radio spectral index, we
confirm that the Eye contains ionized gas, but that this is embedded
within a dense molecular core.  The spectral energy distribution can
be modelled as a two-component blackbody + greybody, peaking at far--IR
wavelengths. The knots are UC \hr\ regions, and the core contains a
luminous ($\sim 2 \times 10^4$\,\lsol), embedded, massive young
stellar source.  We also propose a geometrical model for the Eye to
account for both its spectral energy distribution and its morphology.
\end{abstract}

\begin{keywords}
stars: formation -- \hr\ regions -- ISM: individual: Tornado Nebula, Eye
of Tornado, G357.63-0.06, G357.7-0.1  -- ISM: molecules -- supernova remnants
\end{keywords}

\section{Introduction}
\label{sec-intro}
The Tornado Nebula (G357.7-0.1) is an unusual non-thermal
radio source located close, in projection, to the Galactic centre.  It
has been classified as a supernova remnant on the basis of its
filamentary structure, steep radio spectrum and linear polarization
\citep{milne79,clark76,caswell80,shaver85-1}.  
Its unique structure has, however, prompted other
interpretations, such as an extra-galactic head-tail source 
\citep{weiler80,caswell89}, an accretion powered nebula 
\citep{helfand85,becker85} or a precessing twin-jet system
\citep{stewart94}.  The nebula has an axially symmetric
structure, with multiple components.  The bright western part (the
`Head') is connected to a larger filamentary arc in the centre, which
in turn is connected to an extended `Tail' to the east containing
regularly spaced loops \citep[see][]{stewart94}.  The entire structure
extends over 10 arcmin.  X--ray emission from a thermal plasma is
associated with the Head, as well as from regions of the Tornado's
Tail, and led \citet{gaensler03} to suggest that the Tornado is a
`mixed morphology' supernova remnant.

A compact radio source, the `Eye' (G357.63-0.06), is located about 30
arcsec west of the Head.  Its nature is the subject of this paper.
The apparent proximity to the Tornado, and location along its axis of
symmetry, suggests a connection between the two.  For instance, the
Eye has been interpreted as an accreting binary system responsible for
the formation of the Tornado Nebula \citep{becker85}.  It has also
been suggested that it may be a high proper motion pulsar with the
Tornado as its associated SNR
\citep{shull89}.  However, the flat radio spectrum of the
Eye, together with the {\em IRAS} flux densities, strongly suggest
that it is an \hr\ region, rather than a non-thermal source
\citep{shaver85-2} and is possibly quite unrelated to the Tornado. The
X--ray emission from the Head is also clearly not associated with the
Eye \citep{gaensler03}.

An understanding of both the Tornado and the Eye has been hampered by
the debate regarding whether the two objects are related.  In this
paper we present infrared recombination and molecular line data for
the Eye that allows its kinematic distance to be determined.  It
places the Eye at the same distance as the Galactic centre, and in
front of the Tornado Nebula.  The Eye and the Tornado are therefore
separate sources.  In a related paper \citet{brogan03}, making use of
our near--IR recombination line data to guide their observations with
the VLA, obtained radio recombination line data to reach the same
conclusion. We then proceed to interpret other infrared and radio data
for the Eye in terms of a massive young stellar object in the core of
a compact, dense molecular cloud.  We describe the observations in the
next section, present the results in \S\ref{sec-results}, and discuss
them in \S\ref{sec-discussion}, before drawing conclusions in
\S\ref{sec-conclusions}.

\section{Observations}
\label{sec-obs}
This paper presents observations, taken at infrared, millimetre and
radio wavelengths, of the Eye (G357.63-0.06), a thermal radio source
at the Head of the Tornado Nebula (G357.7-0.1).  The Eye is centred
at $\alpha, \delta=17^{\rm h}~36^{\rm m}~52^{\rm s}$, 
$-30\degr~57\arcmin~22\arcsec$ (B1950).  
The data described here was obtained using a number of
different facilities: the Anglo Australian Telescope (AAT) and UK
Infrared Telescope (UKIRT) for the near-infrared, the Swedish--ESO
Submillimetre Telescope (SEST) for the millimetre and the Very Large
Array (VLA) for the radio data. We also make use of archived data from
the {\em Midcourse Space
Explorer}\footnote{http://irsa.ipac.caltech.edu/ipac/msx/msx.html}
({\em MSX}) and {\em Infrared Astronomy
Satellite}\footnote{http://www.ipac.caltech.edu/ipac/iras/iras.html}
({\em IRAS}) space missions.  We describe how these data were obtained
and processed in the following sub-sections.

\subsection{AAT Observations}
\label{subsec-aat}
The 3.9\,m Anglo Australian Telescope\footnote{The Anglo Australian
Telescope is a bi-national facility operated by the Anglo Australian
Observatory.}  was used with the IRIS near-infrared camera
\citep{allen93} on 14--15 July 1994 to obtain $J$, $H$ and $K_n$
(i.e. 1.25, 1.65 and 2.15\microm\ respectively) broad-band continuum
images, and line images through narrow-band (1\% width) filters
centred at the wavelengths of the hydrogen \brg\ (2.167\microm)
recombination line, and the H$_2$ 2.122\microm\ line.  The images
included both the Eye and portions of the Tornado Nebula.  Only some
of these data are presented here, showing the Eye, and were obtained
with 0.61 arcsec pixel scale and 80 arcsec field of view.  The final
images were created by combining 9 overlapping frames, each with 60 s
integration time, to cover an area roughly $2.5 \times 2.5$ square
arcmin in size.  `Standard' reduction methods were used, involving
linearisation, flat-fielding with a dome flat, sky-subtraction using
the median of the 9 frames, and cleaning of bad pixels.  The separate
frames were then combined using appropriately determined pixel
offsets, to produce a single image.  A coordinate frame was applied by
registering the image with a (lower sensitivity) 2.2\microm\ band
image obtained from the Two Micron All Sky
Survey\footnote{http://www.ipac.caltech.edu/2mass/} (2MASS) data base.
The images were flux calibrated through measurements made of the
standard star HD 159402 (with a $K$--band magnitude of 8.140) taken
through the same filters as the source data.

\begin{figure*}
\centering
\hspace*{-1cm}
\includegraphics[height=12cm]{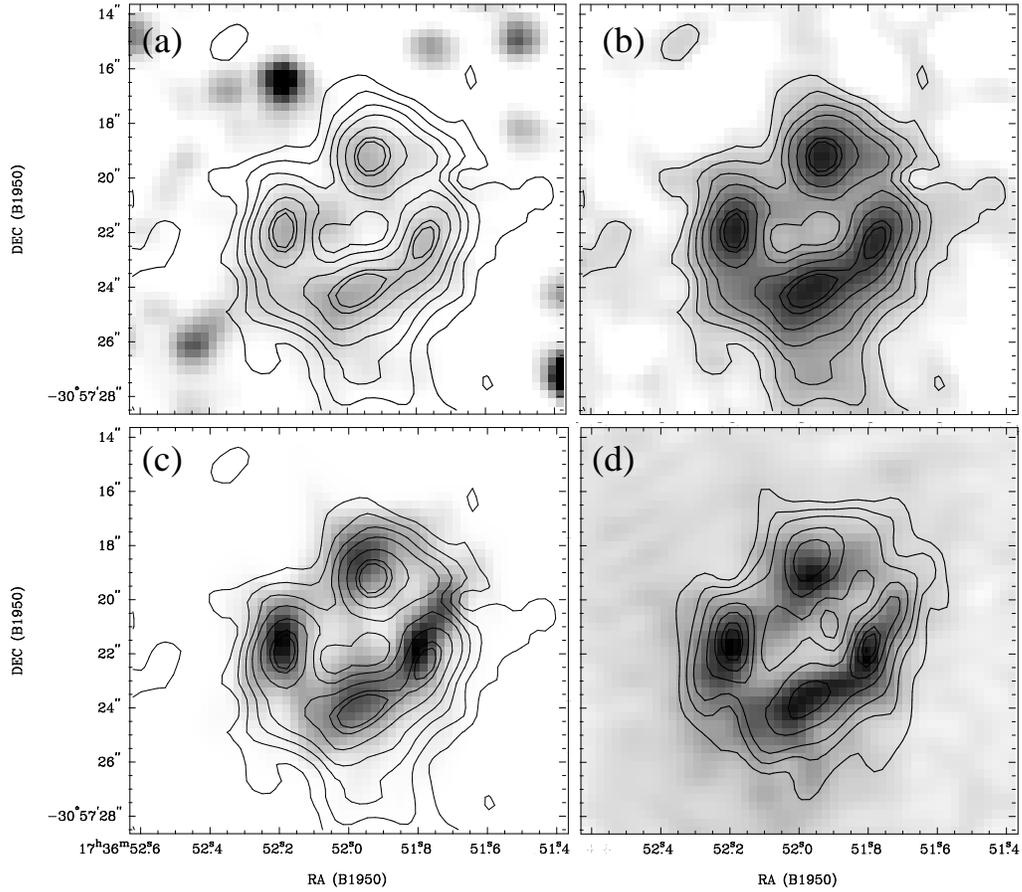}
\vspace*{0cm}\caption{Infrared and radio images of the Eye. 
(a) 2.17\microm\ continuum image
of the field centred on the Eye, overlaid with the contours of
continuum subtracted \brg\ line emission; (b) Continuum subtracted
\brg\ 2.167\microm\ line image of the Eye, overlaid with contours of
the line intensity.  Contours are at 4.8, 8.0, 11.2, 14.4, 19.2, 24.0
and 25.6 \ee{-18} W\,m$^{-2}$\,arcsec$^{-2}$; (c) 20\,cm radio
continuum image overlaid with the same contours of \brg\ line emission
and (d) 6 \,cm radio image overlaid with contours of 20\,cm radio
emission. The 6\,cm greyscale has a range from $-0.4$ to 3.3
mJy\,beam$^{-1}$ and the 20\,cm contours are at 8, 26, 52, 104, 156,
208, 234 and 258 $\times 10^{-3}$ mJy\,beam$^{-1}$.  The FWHM beam
sizes are $2\farcs 7 \times 1\farcs 1$ and $1\farcs 2 \times 0\farcs
9$ at 20 and 6\,cm, respectively. Coordinates are in B1950, and the
frames are the same size in all panels. The IR data was obtained on
the AAT and the radio data from the VLA\@.}
\label{fig-comp}
\end{figure*}

Figure~\ref{fig-comp}a shows the continuum image obtained in the
2.17\microm\ filter in the field around the Eye.  Aside from the small
region of nebulosity associated with Eye, all that the infrared images
show are the crowded star fields apparent in any view directed towards
the central regions of the Galaxy.  Optical images of the same fields,
for instance the Digital Sky
Survey\footnote{http://archive.stsci.edu/dss/} (DSS), show only a
small fraction of the stars seen at 2\microm, the result of extinction
from foreground dust.  While we do not show the results here, an
analysis of the $JHK$ photometry of the field \citep{hoang95} shows
that the stars are typically $\sim$1.5 magnitudes fainter at $H$, and
$\sim$3 magnitudes fainter at $J$, than they are at $K$--band.
Plotting their positions on an infrared colour-colour diagram
indicates a typical visual extinction of A$_{\rm V} \sim 20$ mag.\ for
the stars in the field.

Morphologically, the images of the Eye at 2.12\microm\ and
2.17\microm, as well as in the broad-band $K_n$ filter, are very
similar. A 1.5--2.5\microm\ spectrum, also obtained with IRIS (not
shown), shows that the emission in this band is dominated by the \brg\
line and the \hei\ 2.058\microm\ recombination line, with a weak
continuum.  Therefore, the emission seen through the 2.12\microm\
filter image can be attributed to continuum (and not to \h\ line
emission), and may be used to subtract the continuum from the
2.17\microm\ data to yield a pure \brg\ line image.  This was done by first
registering the two images, and lightly smoothing to ensure the
effective seeing was equivalent in both.  A scaling factor was
determined that minimised the residuals for the stars when the two
images were subtracted.  The \brg\ line image is shown in
Figure~\ref{fig-comp}b, overlaid with contours of the line intensity.
The Eye is not seen in either the $J$ or $H$ band images obtained with
IRIS, which yields an upper limit on the intensity of any source to 19
and 18 magnitudes in these bands, respectively.

\subsection{UKIRT Observations}
\label{subsec-ukirt}

The 3.8 m UK Infrared Telescope was used with the CGS4 spectrometer
 \citep{mountain90} on 13 August 1997 to measure the line profile of \brg\  
2.167\microm\ and the \hei\ 2$^1$P--2$^1$S 2.058\microm\ recombination lines.
The echelle mode of the spectrometer was used, providing a spectral
resolution of 16\kms\ and bandpass of 1,800\kms.  A long slit, of
width 1.2 arcsec, was placed through the southern and the western knots of
the Eye (i.e. orientated 57\degr\ W of N), and the telescope nodded 
30 arcsec along
the slit direction.  1600 s of integration time was obtained on both
lines.  Data reduction consisted of extracting the data from the rows
of the array where the source appeared, subtracting the sky emission,
and wavelength calibrating the spectrum using both an argon arc lamp and
OH airglow lines in the data.  Dividing by the result of equivalent
steps made on measurements of the standard star (BS 6522) both
corrected for sensitivity variations with wavelength, and provided a
flux calibration.

\begin{figure}
\centering
\includegraphics[height=12cm]{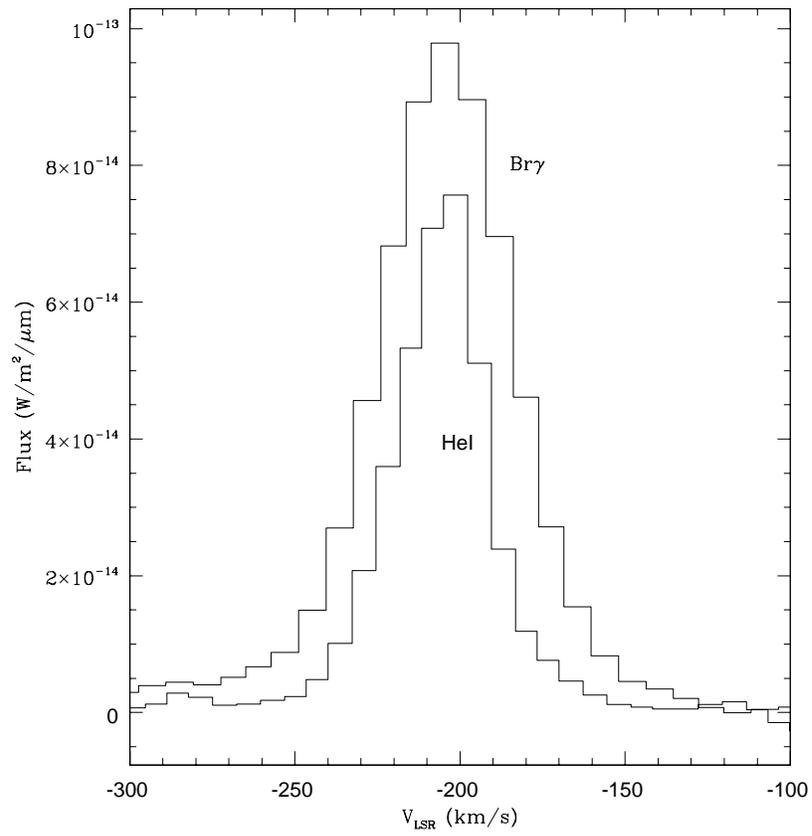}
\caption{Profiles of the \brg\ 2.167\microm\ and \hei\ 2.058\microm\ emission
lines from the Eye, obtained with the UKIRT, with the wavelength scale
set to the velocity, in\kms, measured with respect to the local
standard of rest.  The flux scale is in W\,m$^{-2}$\,\microm$^{-1}$.}
\label{fig-brg}
\end{figure}

The profiles of the two lines are shown in Figure~\ref{fig-brg}, with
the velocity scale set to the local standard of rest (LSR).  The total
flux measured for the \brg\ line is about half that determined through
imaging the line, but this discrepancy is not surprising since the
slit only measures emission from two of the knots.  Not apparent in
this figure are several additional emission features with \VLSR\
velocities from $-$300\kms\ to $-$900\kms\ in the \brg\ line spectrum,
that are 5--10\% of its strength.  These are not, however, high
velocity features associated with the
\brg\ line, but are \hei\ (n=7--4) transitions \citep[see][]{najarro94}.

\subsection{SEST Observations}
\label{subsec-sest}

\begin{figure}
\centering
\rotatebox[origin=cc]{-90}{\includegraphics[height=12cm]{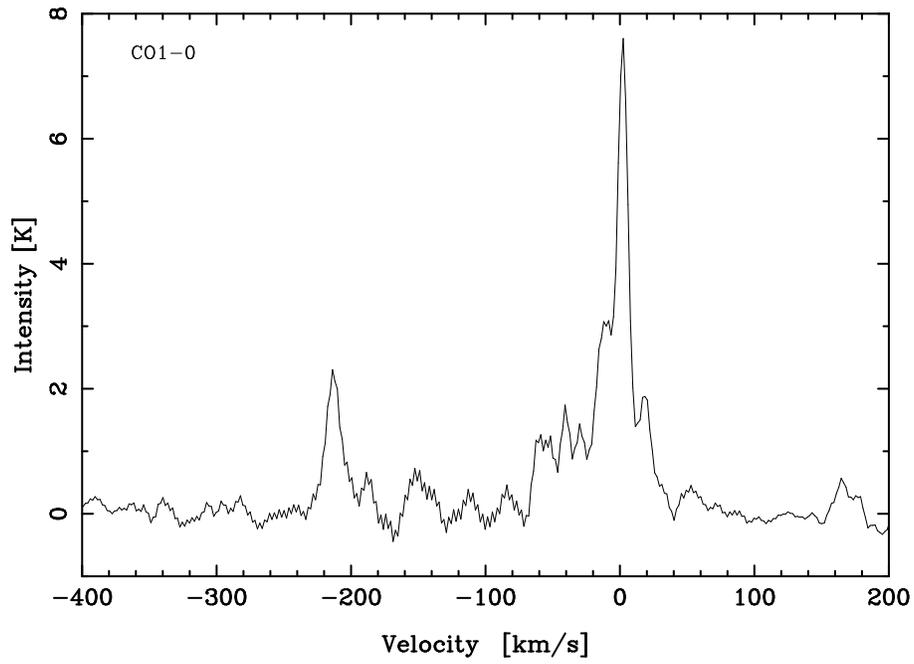}}
\caption{Profile of the \co\ J=1--0 115\,GHz line towards the Eye, obtained
with the SEST at 45 arcsec angular resolution, with velocity measured
with respect to the local standard of rest.  The feature at
$-$205\kms\ is associated with the Eye, as discussed in the text. The
intensity is in units of the main beam brightness temperature.}
\label{fig-co}
\end{figure}

The 15\,m Swedish--ESO Submillimetre Telescope (SEST\footnote{The
Swedish--ESO Submillimetre Telescope is operated by the Swedish
National Facility for Radio Astronomy, the Onsala Space Observatory,
and by the European Southern Observatory.}) was used on 6 June 2000 to
obtain a profile of the \co\ J=1--0 line emission at 2.6 mm from the
Eye.  The angular resolution of these data is 45 arcsec, with 1.5\kms\
spectral resolution (after smoothing), as part of a larger
investigation associated with a study of \oh\ maser emission from the
Tornado \citep{lazendic03a,lazendic03b}.  Data analysis consisted of
baseline subtraction and Hanning smoothing.  The data were corrected
for atmospheric absorption by measurements of a blackbody calibration
source, and had a main beam efficiency of 0.74 applied, to provide the
brightness temperature on the flux scale.  The line profile is shown
in Figure~\ref{fig-co}.  The feature at $-$210\kms\ seems to be
confined to a region within $\sim 90$\,arcsec around the Eye, but this
conclusion is drawn from the data sampled with 60 arcsec spacing
\citep[see][]{lazendic03b}. Therefore, observations with better
sampling and angular resolution are needed to map the distribution of
the molecular cloud associated with the Eye.

Observations of the 1.2\,mm continuum emission were also obtained with
the 37-channel SIMBA bolometer array on 4 June 2002.  The FWHM of each
element is $23''$ with the separation between each $44''$. A fast
scanning mode was used (80 arcsec s$^{-1}$) to obtain a map of size
$400'' \times 600''$ (az $\times$ elev).  The map is comprised of 51
sub-scans, each separated by $8''$, and took approximately 15 minutes
to complete.  Three separate maps were obtained and coadded to produce
the final image.  Pointing and sub-reflector focussing were checked
prior to the observations and sky-dip calibrations performed to obtain
the atmospheric optical depth.  The planet Uranus was used for flux
calibration, with an assumed flux density of 40.3 Jy.  No detection of
the Eye was made at 1.2\,mm, with a $3\sigma$ upper limit of 39\,mJy.

\subsection{VLA Observations}
\label{subsec-vla}
Radio continuum observations of the Eye of the Tornado were carried
out at 6 and 20\,cm on 28 April 1985 and 19 July 1991 with the Very
Large Array of the National Radio Astronomy Observatory\footnote{The
National Radio Astronomy Observatory is a facility of the National
Science Foundation, operated under a cooperative agreement by
Associated Universities, Inc.} in its BnA and A array configurations,
respectively.  We obtained the data from the VLA archives. The phase
centre for both observations are at $\alpha, \delta=17^{\rm h} 36^{\rm
m} 52.1^{\rm s}, -30\degr 57\arcmin 21\farcs 6$ (B1950).  Standard
calibration used 1748--253 and 3C286 as the phase and amplitude
calibrators, respectively.  Standard self-calibration procedure was
also applied to each data set.  The CLEANed beams in the final images
are $2\farcs 7 \times 1\farcs 1$ (FWHM) at 20\,cm, and $1\farcs 2
\times 0\farcs 9$ (FWHM) at 6\,cm, with the respective rms noise being
0.2 and 0.13\,mJy\,beam$^{-1}$. The fluxes at 6 and 20\,cm are listed
in Table~\ref{tab-flux}.  Figure~\ref{fig-comp}c and \ref{fig-comp}d
show an overlay of the \brg\ contours on the 20\,cm radio continuum,
and the 20\,cm contours on the 6\,cm image, respectively.  While there
appears to be a small offset in this Figure between the 20\,cm image
and both the 6\,cm and the \brg\ images, of order 0.5 arcsec, we do
not believe this is real as it is much less than the spatial
resolution of the data at 20\,cm.

\subsection{{\em MSX} and {\em IRAS} Data}
\label{subsec-msx}

Infrared continuum fluxes from 8 to 21\microm\ and from 12 to
100\microm\ have been obtained from the {\em MSX} and {\em IRAS} sky
surveys.  The {\em IRAS} fluxes were obtained from the point source
catalogue, whereas the {\em MSX} fluxes were determined from the
calibrated survey images (and not from the published point source
catalogue, which can be in error).  The source is unresolved in both
the {\em MSX} and {\em IRAS} images, the former limiting the source
size to less than 20 arcsec in extent.  Examination of the {\em MSX}
and SIMBA images fails to show any sources which could contribute
significantly to the far--IR fluxes within the {\em IRAS} beam, so we
have attributed all the flux measured by {\em IRAS} as originating
from the Eye. We also note that no source is seen associated with the
Eye in the 2MASS near--IR sky survey.  The infrared fluxes are listed
in Table~\ref{tab-flux}.

\section{Results}
\label{sec-results}

The Eye is resolved by the near--IR and radio measurements as a
compact \hr\ region (see Fig.~\ref{fig-comp}), and therefore must be
undergoing massive star formation.  It consists of four knots of
emission, each about 1.5 arcsec across and of similar brightness,
symmetrically placed about the perimeter of a circle 6 arcsec across.
There are faint extensions extending $\sim$2 arcsec to the south and
to the west in the \brg\ image, but no emission from its centre.

The emission velocity of the \brg\ and CO lines is $-$205\kms.  It is
also close to the emission velocity for the H92$\alpha$ line that has
recently been measured by \citet{brogan03}.  This velocity allows a
kinematic distance to be determined for the source, despite the
proximity of the sight line to the Galactic centre.  The source is
found to be at the same distance as the Galactic centre, and separated
(probably foreground--see \S\ref{subsec-tornado}) from the Tornado.
The IR continuum fluxes can then be used to derive a luminosity for
the Eye.  Modelling its spectrum shows that it is a heavily embedded
massive young stellar object.  The radio and \brg\ fluxes can then be
used to derive an emission measure and an extinction to the Eye, and
to derive the density of the ionized gas in the source.  The knots are
found to be ionized cavities within a dense molecular core which
appears to have only a single massive star forming within it.  The
analysis that has led to these conclusions is discussed in the
following sub-sections.

\subsection{Distance}
\label{subsec-distance}
\subsubsection{The Eye}
Several features are apparent in the CO profile (Figure~\ref{fig-co}),
with a number of features extending from $-$60\kms\ to $+$20\kms, with
a peak at $+$2\kms.  These undoubtedly arise from distinct molecular
clouds at different distances along the sight line towards the central
part of the Galaxy.  Particularly notable, however, is a 10\kms\ wide
feature at \VLSR\ of $-$205\kms.  This velocity is the same as the
peak emission velocities in the hydrogen and helium infrared
recombination lines (Figure~\ref{fig-brg}), allowing us to identify
these three lines as arising from the same source along this sight
line.  Determining a kinematic distance along such a sight line,
within a few degrees of the Galactic centre ($b$=357\fdg 6,
$l$=$-$0\fdg 1), generally cannot be done since the source's motion
will usually largely be tangential to the sight line.  Non-circular
orbital motions then lead to large uncertainties in the derivation of
the distance.  However, for a source whose galactic orbital motion is
mostly directed towards the Earth this uncertainty is greatly
diminished.  Correspondingly, the source must be close to the Galactic
centre, and near to the tangential point of its Galactic orbit, in
order for this to apply.  We derive a kinematic distance of
8.5$\pm$0.1 kpc for the Eye. The angular separation from the Galactic
centre places it $\sim 300$\,pc away from it.

While a kinematic velocity this large is unusual, it is not unique.
In the CO J=1--0 line survey of the inner galaxy \citep{bania80} two
high velocity molecular features, at $l$=357\fdg 5, $b$=0\degr\ with
\VLSR\ of $-$210\kms\ and $-$190\kms\ are seen. The line of sight
velocity of the Galactic rotation curve peaks at $-$240\kms\ at the
same galactic longitude
\citep{burton78}.  Several \hr\ regions have also been identified in
the Sgr E region ($l$=358\fdg 7, $b$=0\degr), about 200 pc from the
Galactic centre, from the  Molonglo Observatory Synthesis Telescope 
(MOST) 843 MHz radio continuum survey
\citep{gray93}.  Radio recombination lines were later measured from
many of these sources, and found to have velocities ranging from
$-$200\kms\ to $-$215\kms\ \citep{cram96}\@. \citet{lockman96} have
also measured radio recombination lines from two compact sources in
the same region and find similar line of sight velocities (at
$l$=358\fdg 797, $b$=+0\fdg 058, \VLSR=$-$207\kms\ and $l$=358\fdg
974, $b$=$-$0\fdg 021,
\VLSR=$-$193\kms).   \citet{caswell87}, in a
tabulation of the H109$\alpha$ and H110$\alpha$ recombination lines,
measured $-$212\kms\ for the \hr\ region G358.623--0.066\@. Therefore,
the identification of the Eye as an
\hr\ region situated at the distance of the Galactic centre, whose
line of sight velocity is largely determined by Galactic rotation,
seems to be a reasonable one.

\subsubsection{The Tornado}
\label{subsec-tornado}
The distance to the Tornado Nebula is not so well determined, but the
evidence suggests that it is further away than the Eye. \hi\ 21 cm
absorption measurements against the Tornado \citep{Radhak72}, show a
feature at $-$61\kms\ associated with the 3\,kpc spiral arm (as well
as features at $-$100, $-$27, $-$10, +7 and +18\kms).  This places it
further than 5\,kpc away. \oh\ maser emission has been detected next
to one edge of the Tornado \citep{frail96,yz99}, with the
interpretation being that this arises from the interaction of a SNR
with a molecular cloud.  This identification of the Tornado with an OH
maser means that the emission velocity, $-$12.4\kms, which is the same
as that of one of the CO features along this sight line, can be used
to derive a kinematic distance, assuming Galactic rotation.  This is
11.8 kpc, taking the far distance, since the near-distance would place
it only a few hundred pc away.  While the accuracy of this assignment
is in some doubt given the proximity of the sight line to the Galactic
centre, the emission velocity is very different to that associated
with the Eye.

\citet{brogan03} also detected a weak \hi\ absorption feature 
at $-$210\kms\ towards the Head, a similar velocity to the molecular
cloud associated with the Eye. The conclusion, therefore, from all
these pieces of evidence is that the Eye is most likely foreground of
the Tornado Nebula, and that they are unrelated objects.  Its striking
appearance at the head of the Tornado is simply a chance projection
along the line of sight.  We therefore consider the Eye and the
Tornado to be unrelated objects, and in the rest of this paper we use
the data gathered on the Eye to interpret this source.

\subsection{Spectral Energy Distribution}
\label{subsec-sed}
\begin{figure}
\centering
\includegraphics[height=12cm]{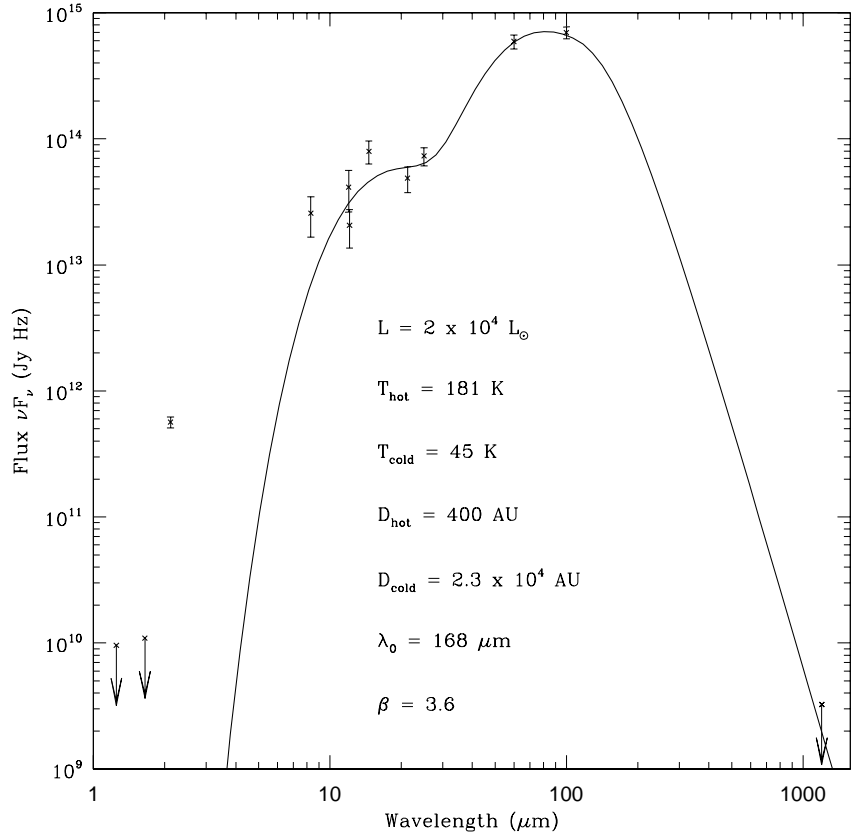}
\caption{Spectral energy distribution for the Eye.  Mid-- and far--IR 
fluxes have been obtained from the {\em MSX} and {\em IRAS} survey
databases, and the near--IR and mm-data from this work, as listed in
Table~\ref{tab-fit}.  Error bars show an estimated absolute error, and
were used to determine a minimum $\chi^2$ fit for a two temperature
greybody model to the data.  The best such fit is shown by the solid
line, with the parameters determined from the fit indicated in the
figure. $\rm D_{hot}$ and $\rm D_{cold}$ are the diameters equivalent
to the angular size of the hot and cold components of the fit, at the
distance to the source. Note that the near--IR data points were not
included in the fit as the 2\microm\ continuum data most likely
represents scattered stellar flux from the edges of the ionized
cavities, and not re-processed thermal radiation from dust. The three
points with arrows show the 1$\sigma$ upper limits on the flux at their
respective wavelengths.}
\label{fig-sed}
\end{figure}

Table~\ref{tab-flux} lists the infrared continuum fluxes for the Eye.
These data can be used to estimate the physical conditions of the
source, such as it temperature, size and luminosity.  We have fitted
the mid--IR, far--IR and mm data with a two-component blackbody +
greybody of the form\footnote{We have not included the near--IR data
in the fit because, since the 2\microm\ continuum is morphologically
similar to the recombination line data, it most likely arises from
scattered flux in, or at the edge of, the ionized cavity and is not
re-processed stellar radiation from the dust.}
\begin{equation}
\Omega_{\rm hot} B_\nu(T_{\rm hot}) + \Omega_{\rm cold} B_\nu(T_{\rm cold}) \epsilon_{\rm dust}.  
\end{equation}
$T_{\rm hot}$ and $T_{\rm cold}$ represent two temperature components
to the dust emission, with $\Omega_{\rm hot}$, $\Omega_{\rm cold}$
being the angular size associated with each component.  The dust
emissivity for the colder component, $\epsilon_{\rm dust}$ (which
dominates the flux at sub--mm and longer wavelengths), is given by ($1
- e^{-\tau}$), where the optical depth is $\tau = \tau_0
(\nu/\nu_0)^{\beta}$, with $\tau_0$ set to 1.  Here $\nu_0$ is the
frequency at which the optical depth is unity.  $\beta$ is the dust
emissivity index, which is uncertain, but thought by most authors to
be in the range 1.5 to 2 \citep[see e.g.][]{chini86,dent98}.  The
parameters for the best fit (minimum $\chi^2$) model, after making
some plausible estimates on the absolute errors of the {\em MSX} and
{\em IRAS} data, are listed in Table~\ref{tab-fit}.  The values,
though, should only be regarded as representative, due to both the
simplicity of the model and the uncertainties inherent the data.  The
data are consistent with the luminosity of the Eye being produced by a
warm ($\sim$180 K), unresolved ($\sim$0.05 arcsec diameter $\equiv
400$\,AU @ 8,500\,pc) blackbody source, at the core of an extended
($\sim$2.7 arcsec $\equiv 23,000$\,AU $\equiv 0.1$\,pc), cold
($\sim$45 K), greybody.  The emission is optically thick for
wavelengths shorter than $\sim$168\microm.  The spectral energy
distribution (SED) for the best fit is shown in Figure~\ref{fig-sed}.
We also note that the best fit value determined for the angular size
is reasonably close to that measured in the
\brg\ and radio images.  Since the fit was made to the (unresolved)
{\em MSX} and {\em IRAS} data, this gives some confidence that the
model is indeed providing a realistic parameterization of the source.
Integrating under the best fit SED gives a total flux of 8.9\ee{-12}
W\,m$^{-2}$, which, at a distance of 8.5 kpc, implies a source
luminosity of $\sim$2.0\ee{4}\lsol.  The cold component dominates the
luminosity, providing an order of magnitude more flux than the warmer
component.  This luminosity is similar to that expected from a
B0--B0.5 main sequence star \citep{panagia73}, whose mass is
$\sim$13\msol.

The dust emissivity index, $\beta$, is poorly constrained by the data,
with a formal fitted value of 3.6.  The lack of detection of the Eye
at 1.2\,mm, however, implies a particularly steep index, considerably
in excess of the $\beta = 2$ typically assumed for grains in dense
cores at sub--mm and longer wavelengths \citep[e.g.][]{ossenkopf94}.
However values for $\beta$ of $\sim 2.4$ have recently been reported
for the Galactic centre region \citep{pierce00} and of $\beta \sim
2.5$ in Orion \citep{lis98}, so a larger value than 2 is not
unexpected.  Such high values are generally interpreted as indicating
the presence of thick ice mantles around the grains.  Whether the
emissivity index is as high as 3.6 or not, however, awaits further
data at sub--mm and mm wavelengths.  It would be premature to use the
current data to attempt to constrain the nature of the dust grains in
the Eye further.

The SED also allows a crude calculation of the total gas mass to be
made.  The fit at 185\microm, just longward of where the dust becomes
optically thin, yields 90\,Jy.  At this wavelength \citet{ossenkopf94}
calculate a dust opacity $\rm \kappa \sim 4.2 \, cm^2 \, g^{-1}$ for
grains with thick ice mantles in dense gas, which implies a gas mass
$\rm \sim 80 M_{\odot}$ (note that we have not used the opacity at
1.2\,mm, although this is more normally used to calculate dust masses,
because of the uncertainty in the value for the emissivity index,
discussed above).  This mass is significantly in excess of the stellar
mass derived from the luminosity of the source.  Furthermore, assuming
that the angular size derived for the source (2.7 arcsec) also
provides an appropriate scale length for the Eye, then it implies an
average density for the gas of $\sim$2\ee{6}\cm{-3}.  The column
density through the core would then be $\rm N_{H_2} \sim 5 \times
10^{23} \, cm^{-2} \equiv A_v \sim 500$\,mags. The Eye contains a
dense molecular core, within which lies an embedded, massive young
stellar object.

\subsection{Recombination Lines}
\label{subsec-recomb}
The morphology of the Eye is remarkably similar in the near--IR \brg\
line at 2.17\microm\ and in the radio continuum at 6 cm and 20 cm.
There are four knots, of roughly equal brightness, with emission
extending between them to form the arc of a circle, 6 arcsec in
diameter.  Table~\ref{tab-knots} lists the fluxes from each of the
knots.  There is no emission detected from the centre of the circle.
The nearly equal fluxes at 6 and 20 cm signify a flat-spectrum source,
and thus optically thin thermal bremsstrahlung emission from an \hr\
region \citep[see e.g.][]{rohlfs00}.  \citet{shaver85-2} have previously
reached the same conclusion from lower resolution data.  The radio
emission can thus be used to determine the emission measure for the
source, and the ratio of the radio to the \brg\ line yields the
extinction to it.  The similarity between these wavelengths indicates
that differential extinction within the source is not significant in
determining the \hr\ region morphology.

The 20 cm flux of 62 mJy determines the emission measure to be
1.1\ee{6}\cm{-6} pc, for an assumed \hr\ region temperature of 10,000
K. This implies an electron density $n_e\sim2$\ee{3}\cm{-3}, assuming
the scale length of the emitting region corresponds to the 6 arcsec
size of the Eye (i.e. $\sim$0.2 pc).  If, however, a scale size of
$\sim$1 arcsec is used, as appropriate to the size of the individual
knots, this increases the electron density to $n_e\sim3$\ee{4}\cm{-3}.
Nevertheless, this is still very much less than the $\sim$\e{6}\cm{-3}
density estimated in \S\ref{subsec-sed} for the molecular gas.  The
four emission knots are ionized cavities within the molecular core.

The emission measure determined from the \brg\ line flux,
1.2\ee{5}\cm{-6}\,pc, is an order of magnitude lower than that derived
from the radio.  Assuming this is the result of extinction, as
discussed above, it implies that $\tau_{2.17} \sim 2.2$ (i.e. A$_{\rm
V}\sim 23$ mag).  This value is similar to the average extinction
derived for stars in the field based on their near--IR colours (see
\S\ref{subsec-aat}).  Its level is consistent with that expected
towards the central regions of the Galaxy.  The bulk of this is not
likely to arise from internal extinction within the source.

Correcting for this extinction yields the number of Lyman continuum
photons, N$\rm_{C} \sim 3.7 \times 10^{47}$\,s$^{-1}$, equivalent to a
B0 star \citep{panagia73}.  This is consistent with the spectral type
derived by \cite{brogan03} from their their measurement at 8.3\,GHz,
as well as from the IR luminosity.  If each of the \brg\ knots were
separately produced by an ionizing star, they would only be slightly
less massive, each still being in the range B0--0.5\@.

While we have not imaged the \hei\ 2.058\microm\ recombination line, its
profile is similar to the \brg\ line.  Both lines peak at \VLSR =
$-$205\kms, and are broad; the \brg\ line has a 40\kms\ FWHM and the
\hei\ line has a 30\kms\ FWHM\@.  This is also consistent with the 
measurements of \cite{brogan03} for the 8.3\,GHz H92$\alpha$ line
($V=-210$\kms, $\Delta V = 36 \pm 7$\kms). The thermal line width for
hydrogen at 10,000 K is $\sim$21\kms\ and for helium it is
$\sim$10\kms.  This would suggest that there are non-thermal
velocities of order 30\kms\ in the gas in order to produce the
observed line widths, possibly the result of turbulent flows within
the \hr\ region.  Such line widths are commonly observed in compact
\hr\ regions \citep{garay99}, the line width decreasing as the size
of the \hr\ region increases.  For sizes of order 0.2\,pc turbulent
contributions of $\sim 30$\kms\ are quite typical for compact \hr\
regions, numbers which these measurements of the Eye are consistent
with.  The virial mass derived for the core from these line widths,
$\rm M_{Vir} \sim 210 (\Delta V)^2 R \sim 10^3 \, M_{\odot}$ (with V
in km/s and R in pc; \citet{garay99}) is considerably in excess of the
estimated core mass from the dust emission. However, given these
turbulent velocities, virial equilibrium cannot be expected to hold
and so this value does not provide an estimate for the mass of the
core.

The \hei/\brg\ intensity ratio is $\sim$0.7.  In principle, this ratio
can be used to estimate the effective temperature of the ionizing
star, since it related to the size of the inner He$^+$ zone compared
to the H$^+$ zone within the \hr\ region \citep[e.g.][]{doyon92},
which is directly related to the hardness of the UV spectrum of the
star.  Model calculations show that for a B0 star this ratio would
typically be expected to be $<$ 0.1; i.e. much smaller than the
observed value.  However, in high density \hr\ regions ($n
>$\e{4}\cm{-3}) the 2$^1$P level of \hei\ is not just populated by
recombination reactions of electrons with He$^+$, but can also be
collisionally populated from the 2$^3$S triplet state, invalidating
simple recombination-based calculations of its line strength
\citep[see e.g.][]{lumsden01}.  The \hei/\brg\ line ratio increases
sharply once collisional excitation to the 2$^1$P level of He occurs.
While the line ratio cannot, therefore, be used to quantitatively
constrain the stellar temperature, the measured ratio of 0.7 is
consistent with excitation of the lines in dense gas ($n_e
>$\e{4}\cm{-3}), as has been inferred from its emission measure.

\section{Discussion}
\label{sec-discussion}
The Eye is a dense molecular core within which a massive star is
forming.  Its {\em IRAS} far--IR colours  are log(F$_{100}$/F$_{60}$)=0.3,
log(F$_{60}$/F$_{25}$)=1.3 and log(F$_{25}$/F$_{12}$)=0.6. These values
place are typical of ultra-compact (UC) \hr\ regions
\citep[e.g.][]{wood89}.  The {\em MSX} mid--IR colours 
($\rm F_{21} / F_{8} = 5$, $\rm F_{14} / F_{12} = 4.9$, $\rm F_{14} /
F_{8} = 5.6$) also places the source in the region associated with
massive young stellar objects on the criteria developed by
\citet{lumsden02}, with the colours most similar to those of compact
HII regions in these authors' sample. The Eye harbours a compact \hr\
region, whose ionizing source is also sufficient to account for the IR
luminosity.  The SED is consistent with a massive embedded source, of
luminosity $\sim$2\ee{4} \lsol.  The CO line-width (10\kms) and
inferred size (0.1\,pc diameter) for the Eye also are quite typical of
hot ammonia cores associated with compact \hr\ regions, and
significantly different from what is observed inside dense cores of
giant molecular clouds
\citep[see][]{garay99}.  These parameters for the Eye are
characteristic of a hot molecular core, containing a massive young
stellar object at its centre, and harbouring a developing UC \hr\
region.

Since no central source is visible at 2\microm\ it must be asked
whether each of the four \hr\ knots in fact contains its own central
source, or whether there is a single source powering them all?  It
would appear that the latter is more likely.  Each of the knots has
similar \brg\ and radio continuum fluxes, and they would need to have
similar ionizing fluxes as well as be at much the same evolutionary
stage, if they were to be separate sources.  Four coeval $\sim$B0.5
stars, one in each knot, would seem to be rather less likely than a
single $\sim$B0 star, sited at the centre of the four knots.  The
knots must represent lower density cavities within the molecular core
into which the ionizing flux can penetrate without being quenched.  It
presumably must be a short-lived phase, as the knots will expand
rapidly and eventually merge.

\begin{figure}
\centering
\includegraphics[height=7cm]{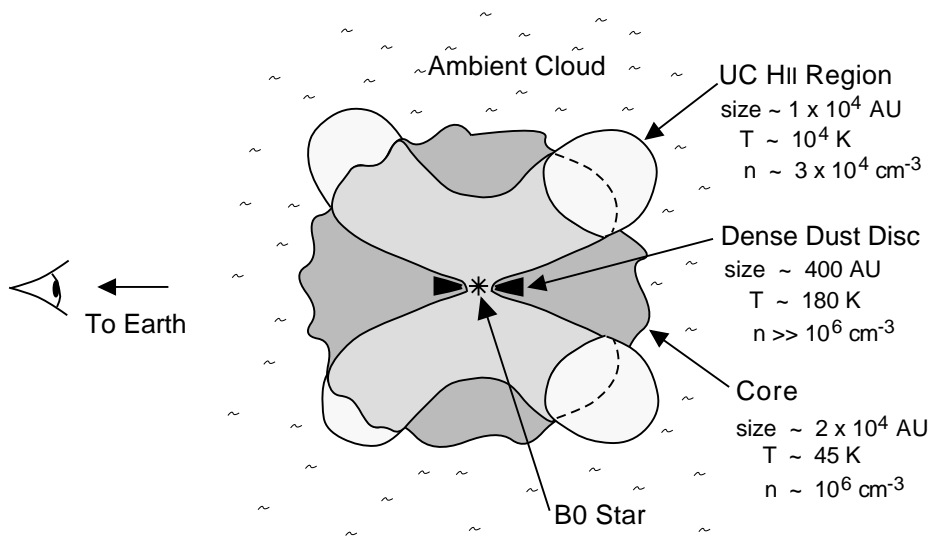}
\caption{A sketch for the model of the Eye that is proposed to
explain both the observed spectral energy distribution and the
morphology, as measured from infrared to radio wavelengths. A young B0
star is obscured from our direct view by a disk-like structure that
extends for $\sim 400$AU around it.  This is heated to $\rm T \sim
180$\,K and intercepts $\sim 10\%$ of the stellar flux.  The
surrounding core, embedded in a lower density molecular cloud, is
about 0.1\,pc across.  It intercepts most of the remaining stellar
radiation, being heated to $\sim 45$\,K in the process.  An \hr\
region is in the process of formation, with the 4 UC \hr\ regions seen
in the \brg\ and radio continuum breaking out of the core in four
quadrants. Inferred physical parameters for the various components are
indicated.}

\label{fig-sketch} 
\end{figure}

The fit to the SED implies that the warm core ($\rm T \sim 180$\,K)
intercepts only around 10\% the flux of the central object.  This
suggests that it either (i) is displaced from the core; (ii) is clumpy
with approximately a 10\% surface filling factor; or (iii) that it has
a non-spherical geometry.  We suggest the latter is the most likely
scenario.  Taking the inferred size from the SED fit, $\sim 400$~AU,
the warm core emission may arise from a flattened, disk-like
structure, blocking our direct view to the central source, but
allowing that source to illuminate the surrounding gas.  The source
heats this gas to $\sim 45$\,K\@.  This produces the far--IR thermal
emission, coming from a cold core, extending out to $\sim 2 \times
10^4$\,AU from the central source.  The core is surrounded by a lower
density molecular cloud, seen in CO line emission. The four nebulae
seen in the \brg\ and radio continuum are ionized cavities, lying
between the disk structure and the inside edge of the molecular cloud,
and are beginning to break out of the cloud.  These cavities have
presumably been created by the expansion of the UC
\hr\ region into the cloud core.  Figure~\ref{fig-sketch} provides a 
sketch of this proposed geometry for the Eye.

The Eye may also be an example of an isolated core where massive star
formation is occurring.  Aside from the measurement of the H92$\alpha$
line \citep{brogan03}, there is no other evidence for star formation
associated with this region. For instance, neither H$_2$O nor CH$_3$OH
maser emission has been reported.  All the activity appears to be
concentrated within a $\rm \sim 0.2$\,pc diameter region, harbouring a
single $\sim$B0 star, whose mass comprises around 10\% that of the gas
in its core.  Of course, there may also be lower mass stars forming
within the core, stars which would still be too faint to be
detected. If so, they would comprise a dense stellar cluster, and are
not part of an extended star forming region.  The Eye appears to
represent an example of an isolated massive star formation, which will
lead to either a single massive star, or a tight stellar cluster whose
luminosity is dominated by a single massive star.


A B0 main sequence star at 8.5 kpc would have a $K$-band magnitude of
11.4, which is more than 5 magnitudes brighter than the upper limit
obtained for the flux of any source at the centre of the Eye.  This
implies that A$_V > 50$ mag, a figure consistent with the column
density derived for the core in \S\ref{subsec-sed}.

\section{Conclusions}
\label{sec-conclusions}

The Eye and the Tornado are two separate sources, with the Eye being
located at the same distance as the Galactic centre, and the Tornado
perhaps 12\,kpc away from the Sun.  Their apparent connection is
simply a chance alignment along our line of sight. The Eye's infrared
luminosity, $\sim$2\ee{4}\lsol, suggests it harbours a massive
protostellar source, perhaps a B0 star ($\sim$13\msol).  It is deeply
embedded within a dense, compact, cold molecular core, whose emission
is dominated by greybody emission from gas at $\sim$45\,K\@.  The core
is 0.1\,pc in diameter, and there is no evidence for star formation
outside it.  A hotter ($\sim 180$\,K), disk-like structure, $\sim
400$\,AU in diameter, may enclose the central source, blocking it from
our direct view. We cannot tell whether lower mass stars are forming
within the core, but nevertheless the Eye appears to represent an
example of isolated massive star formation, with just a single massive
star in the process of formation.

There are several observations that could examine the suggestion of an
isolated massive star formation made here in more detail.
Observations of the sub--mm and mm-wavelength continuum could be used
to verify, or otherwise, the validity of the greybody fit to the
infrared data, and then be used to provide direct constraints on the
mass, of the core.  In particular, they could be used to investigate
whether the steep dust emissivity index inferred here is indeed
correct. Diffraction-limited mid--IR observations with current
8\,m-class telescopes could be used to investigate the nature of the
central source.  Yielding $\sim 0.5''$ spatial resolution, this would
determine whether there is one or more powering sources for the Eye,
and provide better constraints on its SED and evolutionary state.  The
dense molecular core may represent a cold core, soon after it has
started warming up, and in the process of turning into a hot molecular
core \citep[e.g.][]{kurtz00}.  Such cores display a rich chemistry,
and it would be worth targeting the Eye with millimetre and
sub-millimetre spectrometers in order to determine the chemical state
and degree of excitation of its gas.

\section*{Acknowledgments}

We thank Ngoc-Thu Hoang and Ron Stewart for contributing to the early
part of this work, Tom Geballe for explaining what the high-velocity
`features' in the \brg\ profile were, Miller Goss for pointing us to
references which proved invaluable in the interpretation of our data,
and Guido Garay for some insightful discussions about \hr\
regions. Sandy Leggett also obtained the UKIRT data for us, as part of
service observations with that facility.

\clearpage

\begin{table}
\begin{center}
\caption{Integrated Infrared and Radio Fluxes for the Eye. \label{tab-flux}}
\begin{tabular}{@{}lll}
\hline\hline
Wavelength   &  Flux  &  Source \\
\hline
1.25\microm\ & $<$ 1.2\ee{-4} Jy  &  This work, IRIS ($3 \sigma$ limit)\\
1.65\microm\ & $<$ 1.8\ee{-4} Jy  &  This work, IRIS ($3 \sigma$ limit)\\
2.12\microm\ & 4\ee{-3} Jy      &  This work, IRIS \\
\brg\ (2.17\microm) & 6.3\ee{-17} W\,m$^{-2}$ & This work, IRIS \\
8.3\microm\ & 0.7 Jy & {\em MSX} \\
12\microm\  & 1.7 Jy & {\em IRAS} \\
12.1\microm\ & 0.8 Jy & {\em MSX} \\
14.6\microm\ & 3.9 Jy & {\em MSX} \\
21.3\microm\ & 3.5 Jy & {\em MSX} \\
25\microm\ & 6.1 Jy & {\em IRAS} \\
60\microm\ & 119 Jy & {\em IRAS} \\
100\microm\ & 233 Jy & {\em IRAS} \\
1.2mm &	$<$ 3.9\ee{-2} Jy & This work, SIMBA ($3 \sigma$ limit) \\
6 cm  & 7.6\ee{-2} Jy	& This work, VLA \\
20 cm &	6.2\ee{-2} Jy & This work, VLA \\
\hline
\end{tabular}
\end{center}

\medskip
All fluxes are for the continuum, except for the \brg\ line. Note that
the 2.12\microm\ continuum flux, as well as the \brg\ line and radio
continuum fluxes listed, are associated with the \hr\ region and not
with the re-processed stellar radiation from the dust.

\end{table}

\clearpage

\begin{table}
\begin{center}
\caption{Parameters for Best Fit Spectral Energy Distribution to the Eye. \label{tab-fit} }
\begin{tabular}{@{}llll}
\hline\hline
Parameter  &	Description	&	Best Fit Value  &	Range \\
\hline
$T_{\rm hot}$ &	Hot component	& 181 K	& 175 to 187 K \\
$\Omega_{\rm hot}$ & Angular diameter & 0\farcs 047 & 0\farcs 039 to
0\farcs 055 \\
$T_{\rm cold}$ & Cold Component & 44.5 K	 & 43.6 K to 46.2 K \\
$\Omega_{\rm cold}$ & Angular diameter & 2\farcs 7 & 2\farcs 4 to
3\farcs 2 \\
$\nu_0$ & Frequency where $\tau_0 = 1$ & 1.8\ee{12} Hz ($\equiv$ 168\microm) & 1.3\ee{12} to 2.4\ee{12} Hz \\
$\beta$ & Dust emissivity index & 3.6 & 3.0 to $\infty$ \\    
\hline
\end{tabular}
\end{center}

\medskip
Results of fit to the mid--IR, far-IR and mm data, of the form
$\Omega_{\rm hot} B_\nu(T_{\rm hot}) + \Omega_{\rm cold} B_\nu(T_{\rm
cold}) (1 - e^{-\tau})$ with $\tau = \tau_0 (\nu/\nu_0)^{\beta}$,
where $\tau_0$ is set to 1.  $\Omega_{hot}$ and $\Omega_{cold}$ are
determined in steradians, but in the table are quoted with the
equivalent value of their angular diameter for a circular beam.  The
best fit gives the values for the parameters that minimises the
$\chi^2$ difference between the model and the data.  The range gives
the variation that can be made in each parameter, for which the change
in $\chi^2$ from its minimum value is equivalent to a 1$\sigma$
variation, while holding all the other parameters at their best fit
values.  At a distance of 8.5 kpc, 1 arcsec is equivalent to 0.04 pc.

\end{table}

\clearpage

\begin{table}
\begin{center}
\caption{\hr\ Region Emission Fluxes from the Eye. \label{tab-knots} }
\begin{tabular}{@{}llll}
\hline\hline
Knot	& \brg          & 6 cm & 20 cm \\
	& (W\,m$^{-2}$)    & (mJy)  & (mJy) \\
\hline
South	& 1.4\ee{-17}	& 19	& 15 \\
West	& 1.1\ee{-17}	& 13	& 15 \\
North	& 1.2\ee{-17}	& 15	& 15 \\
East	& 1.2\ee{-17}	& 20	& 14 \\
Integrated & 6.3\ee{-17}	& 76	& 62 \\
\hline
\end{tabular}
\end{center}

\medskip
We note that \citet{brogan03} measured 82 mJy and 69 mJy for the
integrated continuum fluxes at 6 cm and 20 cm, whereas
\citet{shaver85-2} obtained 62 and 53 mJy, respectively, at these
wavelengths.

\end{table}


\clearpage

\clearpage

\clearpage

\clearpage

\clearpage

\end{document}